\documentclass[english,prb,footinbib,aps,floatfix,superscriptaddress,tightenlines,twocolumn]{revtex4}
\usepackage[]{fontenc}
\usepackage[latin9]{inputenc}
\usepackage{amsmath}
\usepackage{graphicx}
\usepackage{amssymb}
\usepackage{esint}

\makeatletter
\@ifundefined{textcolor}{}
{%
 \definecolor{BLACK}{gray}{0}
 \definecolor{WHITE}{gray}{1}
 \definecolor{RED}{rgb}{1,0,0}
 \definecolor{GREEN}{rgb}{0,1,0}
 \definecolor{BLUE}{rgb}{0,0,1}
 \definecolor{CYAN}{cmyk}{1,0,0,0}
 \definecolor{MAGENTA}{cmyk}{0,1,0,0}
 \definecolor{YELLOW}{cmyk}{0,0,1,0}
 }

\newcommand{\hc}{\mathrm{h.c.}}
\newcommand{\tr}{\mathrm{tr}}

\newcommand{\1}{\leavevmode{\rm 1\ifmmode\mkern  -4.8mu\else\kern -.3em\fi I}}

\usepackage{times}

\makeatother

\usepackage{babel}

\begin{document}

\title{Optimal quasi-free approximation:\\
reconstructing the  spectrum from ground state energies}

\author{Lorenzo Campos Venuti}

\email{campos@isi.it}

\affiliation{Institute for Scientific Interchange, ISI Foundation, Viale S. Severo
65, I-10133 Torino, Italy}
\begin{abstract}
The sequence of ground state energy density at finite size, $e_{L}$,
provides much more information than usually believed. Having at disposal
$e_{L}$ for short lattice sizes, we show how to re-construct an approximate
quasi-particle dispersion for any interacting model. The accuracy
of this method relies on the best possible quasi-free approximation
to the model, consistent with the observed values of the energy $e_{L}$.
We also provide a simple criterion to assess whether such a quasi-free
approximation is valid. As a side effect, our method is able to assess
whether the nature of the quasi-particles is fermionic or bosonic
together with the effective boundary conditions of the model. When
applied to the spin-1/2 Heisenberg model, the method produces a band
of Fermi quasi-particles very close to the exact one of des Cloizeaux
and Pearson. The method is further tested on a spin-1/2 Heisenberg
model with explicit dimerization and on a spin-1 chain with single
ion anisotropy. A connection with the Riemann Hypothesis is also pointed
out. 
\end{abstract}
\maketitle

\section{introduction}

The way finite-size, thermodynamic, averages approach their infinite
size limit, encodes a wealth of precious information. Take for example
a quantum system at zero temperature, and consider the ground state
energy density at finite size $e_{L}$, $L$ being the linear size
of the system. A piece of common knowledge in condensed matter theory
--not yet a theorem though-- asserts that, for a massive (i.e.~gapped)
system with translation and Left-Right invariance, the approach to
the thermodynamic limit is exponentially fast, that is $e_{L}\rightarrow e_{\infty}+O\left(e^{-L/\xi_{E}}\right)$
where the decay rate $\xi_{E}$ is somehow related to the correlation
length of the system $\xi_{C}$. On the contrary, in a critical theory,
the correlation length is formally infinity and the approach is expected
to be of algebraic type. Moreover for conformal field theories (CFT)
in $\left(1+1\right)$ dimensions and periodic boundary conditions,
it is known that $e_{L}=e_{\infty}-\left(\pi/6\right)cv/L^{2}+o\left(L^{-2}\right)$
\cite{blote86,affleck86} where $v$ is the speed of elementary excitations
and $c$ is the all-important central charge of the CFT. From these
argument it seems clear that it should be possible, by simply looking
at the sequence $e_{L}$, to discern whether the theory is critical
or gapped. In fact, even though the standard method to locate critical
point is that of looking at the gap closure, recent studies have shown
that accurate methods to locate critical points are available, which
resort solely to the computation of ground state averages such as
the total energy and the perturbation \cite{roncaglia07}. 

In this article we will show that the sequence $e_{L}$ provides in
fact much more information. We will show that, by knowing $e_{L}$
for few short sizes $L$, it is possible to reconstruct approximately
the one particle dispersion of the theory. Perhaps most importantly,
analyzing the sequence $e_{L}$, one is able to assess whether the
quasi-particles of the theory have fermionic or bosonic character
together with the effective boundary conditions of the model. 

If the Hamiltonian $H$ under consideration admits a quasi-free representation
(i.e.~the Hamiltonian can be expressed as a quadratic form in Fermi
or Bose operators), our algorithm produces a one-particle dispersion
rapidly converging to the exact one, as we increase the number of
available ground state energies. In practice having 10 energy data
produces a result indistinguishable from the exact one. 

Conversely, if $H$ is truly interacting, we obtain an approximate
one-particle dispersion corresponding to a quasi-free model $\tilde{H}$.
The Hamiltonian $\tilde{H}$ so obtained, is the simplest quasi-free
model whose ground state energies are the same as those observed for
$H$. 

For quasi-free models the mathematical problem is related to that
of reconstructing a function given some of its partial Riemann sums.
As we will show, assessing the convergence speed of this method is
a problem connected to the Riemann hypothesis. Let us then begin by
considering the problem of convergence of Riemann sums.

\section{Convergence of Riemann sums\label{sec:Convergence-of-Riemann}}

As we will show in greater detail in section \ref{sec:Method-and-applications},
the finite-size, ground state energy of any, translation invariant,
quasi-free system, consisting either of bosons or of fermions, is
related to a Riemann sum of the following form\begin{equation}
S_{L}\left(f\right)=\frac{1}{L}\sum_{n=0}^{L-1}f\left(\frac{2\pi n+\vartheta}{L}\right).\label{eq:riemann-sum}\end{equation}
Here $L$ is precisely the size of the one-dimensional lattice %
\footnote{To be precise the natural number $L$ must be intended as the system
length divided by the size of the primitive cell. All lengths scales
to be found in the following are to be intended in units of this lattice
constant. %
} and $f$ a suitable function related to the one-particle dispersion.
The points $k_{n}=\left(2\pi n+\vartheta\right)/L$ are quasimomenta
which define the Brillouin zone $BZ=[0,2\pi)$ %
\footnote{For simplicity of notation and without loss of generality, we defined
the Brillouin zone in $[0,2\pi)$. %
} for finite $L$, and the angle $\vartheta\in\left[0,\pi\right]$
defines general, twisted boundary conditions (TBC). In terms of canonical
operator (either bosons or fermions) $c_{x}$, TBC means $c_{x+L}=e^{i\vartheta}c_{x}$.
The angle $\vartheta$ allows to interpolate continuously between
periodic (PBC) $\vartheta=0$ and antiperiodic (ABC) $\vartheta=\pi$
boundary conditions. 

Since the function $f$ is defined on $[0,2\pi)$ it is useful to
write it as a Fourier series:\begin{equation}
f\left(k\right)=\sum_{n=-\infty}^{\infty}c_{n}e^{ink},\quad c_{n}=\frac{1}{2\pi}\int_{0}^{2\pi}e^{-i\xi n}f\left(\xi\right)d\xi.\label{eq:foureir_series}\end{equation}
In order to obtain an explicit formula for Eq.~(\ref{eq:riemann-sum})
we would like to apply the operator $S_{L}$ to the exponentials appearing
in Eq.~(\ref{eq:foureir_series}) (i.e.~swap $S_{L}$ with the sum).
To this end we need to assume some regularity in the convergence of
the sum (\ref{eq:foureir_series}) such as absolute convergence. Necessary
conditions for absolute convergence of Eq.~(\ref{eq:foureir_series})
are known; for instance it is enough to have $f$, $\alpha$-H\"older
continuous with $\alpha>1/2$, or $f$ of bounded variation and $\alpha>0$.
On physical grounds we can assume such conditions will be satisfied,
they are indeed satisfied in all examples encountered. Now, assuming
absolute convergence of the series in Eq.~(\ref{eq:foureir_series})
we can apply the operator $S_{L}$ to the exponentials obtaining\begin{equation}
S_{L}\left(e^{ink}\right)=\frac{e^{in\vartheta/L}}{L}\frac{\left(1-e^{i2\pi n}\right)}{\left(1-e^{i2\pi n/L}\right)}.\end{equation}
This equation is zero for all natural $n$ not multiple of $L$, whereas
for $n=lL$ with $l\in\mathbb{Z}$, equals $e^{il\vartheta}$. Therefore
we can compactly write $S_{L}\left(e^{ink}\right)=\sum_{l=-\infty}^{\infty}e^{il\vartheta}\delta_{n,lL}$.
Combining this result with Eq.~(\ref{eq:foureir_series}) we obtain
\begin{eqnarray}
S_{L}\left(f\right)-c_{0} & = & \sum_{l=1}^{\infty}e^{il\vartheta}c_{lL}+e^{-il\vartheta}c_{-lL}\label{eq:S_L-theta}\\
 & = & 2\mathrm{Re}\sum_{l=1}^{\infty}e^{il\vartheta}c_{lL},\end{eqnarray}
where the last line follows from the reality of $f$. 

Assume now that the function $f$ is $2\pi$-periodic and \emph{analytic.
}Then there exist a strip in the complex plane $\mathcal{A}=[0,2\pi]\times[-\sigma,\sigma]$
such that $f$ can be extended to an analytic, bounded function on
$\mathcal{A}$. In this case Fourier coefficients can be computed
in the complex plane $z$, integrating on a horizontal line shifted
by an imaginary amount $-i\sigma$. The vertical contributions cancel
each other because of periodicity and we can write\begin{equation}
c_{n}=\frac{1}{2\pi}\int_{0-i\sigma}^{2\pi-i\sigma}e^{-inz}f\left(z\right)dz.\end{equation}

At this point the geometric series of the exponential converges absolutely
and we obtain\begin{equation}
S_{L}\left(f\right)-c_{0}=\mathrm{Re}\left[\frac{1}{\pi}\int_{0-i\sigma}^{2\pi-i\sigma}\frac{f\left(z\right)dz}{e^{i\left(zL-\vartheta\right)}-1}\right].\label{eq:integral_rep}\end{equation}
This formula is remarkable; it allows to show that, for periodic analytic
functions, the convergence of partial Riemann sum to the limiting
integral, is at least exponentially fast in the size $L$. More precisely,
since $1/\left|e^{i\left(zL-\vartheta\right)}-1\right|\le1/\left(e^{\sigma L}-1\right)$
for all $z$ in $\mathcal{A}$ and $\vartheta=0,\pi$, using (\ref{eq:integral_rep})
we obtain \begin{equation}
\left|S_{L}\left(f\right)-c_{0}\right|\le2\max_{z\in\mathcal{A}}\left|f\left(z\right)\right|\frac{1}{e^{\sigma L}-1}.\label{eq:exp_bound}\end{equation}

This shows that, for analytic periodic functions $f$, the partial
sum $S_{L}\left(f\right)$ converges at least exponentially fast in
$L$ to its limiting value. Moreover Eq.~(\ref{eq:exp_bound}) gives
a simple way to compute the decay rate. All we have to do is take
$\sigma$ as large as possible, i.e.~find the largest strip $\mathcal{A}$
where the function is analytic. This amounts to look for the singularity
of $f$ closest to the real axis, let it be $z_{0}$. The imaginary
part $\xi_{E}^{-1}=\mathrm{Im}\left(z_{0}\right)=\max\sigma$ gives
the decay rate and from Eq.~(\ref{eq:exp_bound}) we obtain $\left|S_{L}\left(f\right)-c_{0}\right|\le O\left(e^{-L/\xi_{E}}\right)$. 

As we will discuss in greater detail in section \ref{sec:Method-and-applications},
for quasi-free systems (that is, Hamiltonians quadratic in Fermi or
Bose canonical operators), the ground state energy is exactly given
by an expression of the form of Eq.~(\ref{eq:riemann-sum}) where
the function $f$ is proportional to the one-particle dispersion.
The result above shows that if the one particle band is a positive,
analytic, periodic function, the ground state energy density approaches
its limit exponentially fast. The positivity requirement on the band
implies that the system has a gap. So we reach the conclusion that
for gapped theories with analytic dispersion, the energy density decays
at least exponentailly with the size --at least for the case of quasi-free
systems. Since, from scaling hypothesis, the correlation length is
the only length-scale of the system, we expect the decay rate $\xi_{E}$
to be related to the correlation length. 

Equation (\ref{eq:integral_rep}) cannot be used when the function
is not analytic on $\left[0,2\pi\right]$, in this case the difference
$S_{L}\left(f\right)-c_{0}$ can be estimated using Euler-Maclaurin
formula. In any case, for critical theories where the dispersion vanishes
linearly as $\omega\left(k\right)\sim v\left|k-k_{F}\right|^{z}$
(here $z$ is the dynamical exponent and $k_{F}$ the Fermi momentum)
scaling theory predicts an algebraic approach of the form $S_{L}\left(\omega\right)-c_{0}=O\left(L^{-1-z}\right)$
(see e.g.~\cite{barber_scaling}). An example in this class is given
by the function $\left|\cos\left(k\right)\right|$, which is continuous
but not differentiable. An explicit calculation shows in fact $S_{L}\left(\left|\cos\left(k\right)\right|\right)-2/\pi=O\left(L^{-2}\right)$
consistent with $z=1$. To complete the scenario we should mention
another possibility which gives rise to an algebraic approach to the
thermodynamic limit. Namely the function itself might have a jump
at some point as it happens, for instance, to the function $k$. In
fact $0$ and $2\pi$ should be identified in the Brillouin zone and
this function has a jump of value $2\pi$. In fact in this case we
have $S_{L}\left(k\right)=\pi-\pi/L$. This situation typically takes
place in physical systems when Left-Right symmetry is explicitly broken. 

Our aim here is to reconstruct $f$ given $S_{L}\left(f\right)$,
that is, we would like somehow to invert Eq.~(\ref{eq:integral_rep}).
To this end it is better to do a step back.

\section{Solution of the inverse problem}

Even in the most favorable case, we can hope to have at disposal only
a finite number of partial sum $S_{m}\left(f\right)$, say the first
$L$ (obtained, for instance by Lanczos diagonalization, see sec.~\ref{sec:Method-and-applications}).
This means we can only reconstruct $f$ through $L$ coefficients.
A sensible choice, and in some sense the best one, is that of reconstructing
the first $L$ \emph{Fourier }coefficients of $f$. 

In the rest of the article we will concentrate on the physically most
important periodic and anti-periodic BC's for which $\vartheta=0,\,\pi$.
In these cases the partial Riemann sums are identically zero for $f$
odd, i.e.~$S_{L}^{\left(\vartheta\right)}\left(f\right)=0$ for $f\left(k\right)=-f\left(-k\right)[=-f\left(2\pi-k\right)]$.
Clearly, having at disposal the numbers $S_{L}^{\left(\vartheta\right)}\left(f\right)$
we can only hope to re-construct the even part of $f$. Fortunately,
on physical grounds we can suppose that the function $f$ will indeed
be even in $k$. This corresponds to Left-Right symmetry on top of
translational symmetry and it is a reasonable assumptions valid in
most physical situations. A breaking of Left-Right symmetry happens,
for instance, when inserting a periodic spin chain in a magnetic field
of flux $\vartheta\hbar c/e$ \cite{byers61}. Such a problem can
usually be reformulated into the same problem without magnetic field
and TBC specified by the angle $\vartheta$. All in all considering
$\vartheta=0,\,\pi$ is consistent with assuming $f$ even. 

Assume then that the function $f$ is even so that its Fourier series
has only cosine coefficients. For PBC and ABC, $e^{i\vartheta}$ is
real and, defining $R_{L}\left(f\right)=S_{L}\left(f\right)-c_{0}$,
we can re-write Eq.~(\ref{eq:S_L-theta}) in matrix notation as $\boldsymbol{R}=\mathcal{G}\boldsymbol{a}$,
where boldface indicates column vector and the matrix $\mathcal{G}$
has components given by\begin{equation}
\mathcal{G}_{M,m}=\sum_{l=1}^{\infty}e^{i\vartheta l}\delta_{m,lM}\,.\label{eq:G_matrix}\end{equation}
The components of the vector $\boldsymbol{a}$ are the cosine Fourier
coefficients \begin{equation}
a_{n}=2\mathrm{Re}c_{n}=\frac{1}{\pi}\int_{0}^{2\pi}\cos\left(\xi n\right)f\left(\xi\right)d\xi.\end{equation}
Note that both $\mathcal{G}$ and $\boldsymbol{a}$ are real. To solve
for the first $L$ Fourier coefficients we truncate the equation to
the first $L$ terms to obtain $\boldsymbol{R}^{\left(L\right)}=\mathcal{G}^{\left(L\right)}\boldsymbol{a}^{\left(L\right)}$,
where $\boldsymbol{R}^{\left(L\right)}=\left(R_{1},R_{2},\ldots,R_{L}\right)^{\top}$,
$\boldsymbol{a}^{\left(L\right)}=\left(a_{1},a_{2},\ldots,a_{L}\right)^{\top}$
and $\mathcal{G}^{\left(L\right)}$ is the $L\times L$ matrix with
entries given by Eq.~(\ref{eq:G_matrix}) for $M,m\le L$. For example,
for $L=6$ , the matrix $\mathcal{G}^{\left(6\right)}$ reads\begin{equation}
\mathcal{G}^{\left(L=6\right)}=\left(\begin{array}{cccccc}
q & q^{2} & q^{3} & q^{4} & q^{5} & q^{6}\\
0 & q & 0 & q^{2} & 0 & q^{3}\\
0 & 0 & q & 0 & 0 & q^{2}\\
0 & 0 & 0 & q & 0 & 0\\
0 & 0 & 0 & 0 & q & 0\\
0 & 0 & 0 & 0 & 0 & q\end{array}\right).\end{equation}
with $q=e^{i\vartheta}=\pm1$. Since $\det\left(\mathcal{G}^{\left(L\right)}\right)=q^{L}\neq0$
the matrix is invertible and defines a bijection. It is then possible
to obtain the first $L$ Fourier components approximately via $\boldsymbol{a}^{\left(L\right)}=\left[\mathcal{G}^{\left(L\right)}\right]^{-1}\boldsymbol{R}^{\left(L\right)}$.
One can show that the inverse matrix has the same structure as $\mathcal{G}^{\left(L\right)}$
in the sense that $\left[\mathcal{G}^{\left(L\right)}\right]^{-1}$
has non zero entries in the same places as $\mathcal{G}^{\left(L\right)}$.
More precisely $\left(\left[\mathcal{G}^{\left(L\right)}\right]^{-1}\right)_{i,j}=b\left(j/i\right)$
if $j/i\in\mathbb{N}$ and zero otherwise, for some numeric function
$b\left(n\right)$. Using a similar notation as before, this means
\begin{equation}
\left(\left[\mathcal{G}^{\left(L\right)}\right]^{-1}\right)_{i,j}=\sum_{l=1}^{\left[L/i\right]}b\left(l\right)\delta_{j,li}\,.\end{equation}
Imposing $\left[\mathcal{G}^{\left(L\right)}\right]^{-1}\mathcal{G}^{\left(L\right)}=\1$
we get the equation defining $b\left(m\right)$:\begin{equation}
\sum_{\stackrel{m=1}{j/m\in\mathbb{N}}}^{j}e^{i\vartheta j/m}b\left(m\right)=\delta_{1,j}\,.\label{eq:defining_b}\end{equation}

The above sum extends over all positive divisors of $m$ and is usually
denoted by $\sum_{j|m}$ (read $j$ divides $m$) in the mathematical
literature. Equation (\ref{eq:defining_b}) provides a recursive solution
for $b\left(m\right)$: $b\left(1\right)=e^{-i\vartheta}$ while $b\left(j\right)=-e^{-i\vartheta}\sum_{\stackrel{m=1}{j/m\in\mathbb{N}}}^{j-1}e^{i\vartheta j/m}b\left(m\right)$
for $j\ge2$. Since Eq.~(\ref{eq:defining_b}) is independent of
$L$, it implies that $\left[\mathcal{G}^{\left(L\right)}\right]^{-1}$
is the first $L\times L$ sub-matrix on the diagonal of $\left[\mathcal{G}^{\left(M\right)}\right]^{-1}$
for any $L<M$. In particular any $\left[\mathcal{G}^{\left(L\right)}\right]^{-1}$
can be obtained from the infinite case $\mathcal{G}^{-1}$ and its
entries do not depend on $L$. Now the inverse formula $\boldsymbol{a}^{\left(L\right)}=\left[\mathcal{G}^{\left(L\right)}\right]^{-1}\boldsymbol{R}^{\left(L\right)}$,
in components reads\begin{equation}
a_{k}^{\left(L\right)}\left(f\right)=\sum_{n=1}^{\left[L/k\right]}b\left(n\right)R_{nk}\left(f\right),\label{eq:finite_inverse}\end{equation}
where we explicitly indicated the dependence on the function $f$.
The superscript $\left(L\right)$ indicates that the Fourier coefficients
are obtained only approximately, but with increasing precision the
larger the $L$. The result of the reconstruction is optimal in the
sense that it produces the \emph{unique} trigonometric polynomial
of degree $L$ consistent with the {}``observed'' data $S_{L}$.
In the limit $L\to\infty$ -- that means we know $R_{L}\left(f\right)$
for arbitrary $L$ -- we can reproduce the function exactly and $a_{k}^{\left(L\right)}\left(f\right)\to a_{k}\left(f\right)$.
To be mathematically precise, this last assertion is satisfied provided
$f$ is not too pathological. On physical ground we can safely discard
such pathological cases. On the contrary Eq.~(\ref{eq:finite_inverse})
can be used to obtain information on the function $f$. For what discussed
in the previous section, the only potentially dangerous case is that
of a critical point. Since physically we will identify $R_{L}\left(f\right)$
with the finite size energy density, scaling arguments predict that,
at a critical point, $R_{L}\left(f\right)\sim L^{-1-z}$ where $z$
is the dynamical critical exponent. Now one can use Eq.~(\ref{eq:finite_inverse})
to obtain $a_{k}\sim k^{-1-z}$ which gives rise to a function with
absolutely convergent Fourier series. 

The PBC case $\vartheta=0$ provides some interesting connections
to number theory. In this case in fact Eq.~(\ref{eq:defining_b})
becomes the equation defining the M\"obius function $\mu\left(m\right)$
\footnote{The M\"obius function is defined as $\mu\left(n\right)=\left(-1\right)^{r}$
if $n$ is a product of $r$ distinct primes, while in all other cases
where $n$ contains a square $\mu\left(n\right)=0$%
}, i.e.~Eq.~(\ref{eq:defining_b}) is solved by $b\left(n\right)=\mu\left(n\right)$.
Note also that, for $q=1$, $\mathcal{G}^{\left(L\right)}$ is very
similar to the Redheffer matrix $\mathcal{R}^{\left(L\right)}$ known
in number theory \cite{redheffer77}. The matrix $\mathcal{R}^{\left(L\right)}$
is the same as $\mathcal{G}^{\left(L\right)}$ except for the first
column which is made of one. The importance of the Redheffer matrix
originates from the fact the $\det\mathcal{R}^{\left(L\right)}=M(L):=\sum_{n=1}^{L}\mu\left(n\right)$
where $M\left(L\right)$ is the Mertens function. The statement $M\left(L\right)=O\left(L^{1/2+\epsilon}\right)$
is equivalent to the Riemann hypothesis.

\subsection{Convergence rate\label{sub:Covergence-rate}}

Since we are assuming the function we seek is even, Eq.~(\ref{eq:foureir_series})
becomes $f\left(k\right)=c_{0}+\sum_{n=1}^{\infty}a_{n}\cos\left(nk\right)$.
If we have access to the partial Riemann sums up to $L$ (and the
limiting value $c_{0}$), we can reconstruct an approximate function
given by\begin{equation}
f_{L}\left(k\right)=c_{0}+\sum_{n=1}^{L}a_{n}^{\left(L\right)}\cos\left(nk\right),\label{eq:f_approx}\end{equation}
with $a_{n}^{\left(L\right)}$ given by equation (\ref{eq:finite_inverse}).
We can now ask how fast this method allows to reproduce the function
$f$. The question of convergence of Fourier series has engaged mathematicians
for centuries. The problem of convergence of the series (\ref{eq:f_approx})
is likely to be more complex. Here we will content to give some arguments
for the physically important cases related to massive and critical
theory. 

To be specific we will consider PBC. We can safely assume $f$ to
be square summable. The reconstructed function $f_{L}$ is obviously
also square summable, being a trigonometric polynomial. The $L^{2}\left(\left[0,2\pi\right]\right)$
square distance reads\begin{equation}
\left\Vert f_{L}-f\right\Vert _{2}^{2}=\pi\sum_{n=1}^{L}\left(a_{n}^{\left(L\right)}-a_{n}\right)^{2}+\pi\sum_{n=L+1}^{\infty}a_{n}^{2}.\label{eq:norm2}\end{equation}

Consider first the case where $f$ is periodic and analytic in $\left[0,2\pi\right]$.
In this case a saddle point argument shows that the Fourier coefficients
decay exponentially $c_{n}=O\left(e^{-n/\xi_{F}}\right)$. Moreover
the correlation length $\xi_{F}$ is precisely the same as that appearing
in Sec.~\ref{sec:Convergence-of-Riemann}: $\xi_{F}=\xi_{E}$. Hence
the second sum in Eq.~(\ref{eq:norm2}) is of the order of $e^{-2L/\xi_{E}}$.
To estimate the first sum consider \begin{equation}
a_{n}^{\left(L\right)}-a_{n}=\sum_{m=\left[L/n\right]+1}^{\infty}b\left(m\right)R_{nm}\,.\label{eq:first_term}\end{equation}
For what discussed in Sec.~\ref{sec:Convergence-of-Riemann}, $R_{m}=O\left(e^{-m/\xi_{E}}\right)$.
For PBC $\left|b\left(m\right)\right|=\left|\mu\left(m\right)\right|\le1$
so that $\left|a_{n}^{\left(L\right)}-a_{n}\right|\le\sum_{m=\left[L/n\right]+1}^{\infty}\left|R_{nm}\right|$.
This  implies that, for sufficiently large $L$, $\left|a_{n}^{\left(L\right)}-a_{n}\right|\le O\left(e^{-L/\xi_{E}}\right)$,
so that, in turn, the first sum in Eq.~(\ref{eq:norm2}) is bounded
by $Le^{-2L/\xi_{E}}$. All in all, if $f$ is periodic and analytic
in $\left[0,2\pi\right]$, $f_{L}$ is exponentially close to $f$,
in sense that $\left\Vert f_{L}-f\right\Vert _{2}^{2}\le O\left(Le^{-2L/\xi_{E}}\right)$. 

Consider now the critical case. From a physical point of view, a critical
theory with dynamical exponent $z$ corresponds to excitations vanishing
as $f\left(k\right)\sim k^{z}$, with $z>0$. A scaling argument now
implies that the Fourier coefficients of such a function scale as
$c_{n}\sim n^{-1-z}$. More precisely assume that, for large enough
$n$ $\left|c_{n}\right|\le An^{-1-z}$ with $A$ positive constant.
Let us first consider the second sum in Eq.~(\ref{eq:norm2})\begin{eqnarray}
\left|\sum_{n=L+1}^{\infty}a_{n}^{2}\right| & \le & A^{2}\sum_{n=L+1}^{\infty}n^{-2-2z}\,.\end{eqnarray}
We can use the asymptotic behavior $\sum_{n=M}^{\infty}n^{-\alpha}\sim\left(\alpha-1\right)^{-1}/M^{\alpha-1}$
to estimate $\left|\sum_{n=L+1}^{\infty}a_{n}^{2}\right|\lesssim A^{2}\left(1+2z\right)^{-1}/L^{1+2z}$.
To obtain the behavior of the first sum in Eq.~(\ref{eq:norm2})
we first look at Eq.~(\ref{eq:S_L-theta}) and obtain, for $L$ large
enough $\left|R_{L}\right|\le2AL^{-1-z}\zeta\left(1+z\right)$. Then
\begin{equation}
\left|a_{n}^{\left(L\right)}-a_{n}\right|\le\frac{2A\zeta\left(1+z\right)}{n^{1+z}}\sum_{m=\left[L/n\right]+1}^{\infty}\frac{1}{m^{1+z}}.\end{equation}
Using the same estimate as before we obtain $\left|a_{n}^{\left(L\right)}-a_{n}\right|\lesssim2A\zeta\left(1+z\right)z^{-1}/(nL^{z})$
and so $\sum_{n=1}^{L}\left|a_{n}^{\left(L\right)}-a_{n}\right|^{2}\lesssim A'/(L^{2z})\sum_{n=1}^{L}n^{-2}\lesssim A''/(L^{2z})$.
So it seems that the convergence rate of $\left\Vert f_{L}-f\right\Vert _{2}^{2}$
is of the order of $L^{-2z}.$ Here we would like to conjecture that
the first sum in Eq.~(\ref{eq:norm2}) actually introduces corrections
that are roughly of the same order as the second, i.e.~$\left\Vert f_{L}-f\right\Vert _{2}^{2}\sim L^{-1-2z}$.
The argument is based on the Riemann hypothesis. Consider again $a_{n}^{L}-a_{n}$
in the PBC case. For $L$ sufficiently large we have \begin{equation}
a_{n}^{L}-a_{n}\sim n^{-1-\zeta}\sum_{m=\left[L/n\right]+1}^{\infty}\mu\left(m\right)m^{-1-\zeta}.\label{eq:coeff_mu}\end{equation}
An equivalent statement to the Riemann hypothesis is that the Mertens
function $M\left(x\right):=\sum_{m=1}^{x}\mu\left(m\right)$ satisfies
$M\left(x\right)=O\left(x^{1/2+\epsilon}\right)$ for any positive
$\epsilon$. Using partial summation %
\footnote{$\sum_{k=M}^{N}f_{k}\Delta g_{k}=f_{N+1}g_{N+1}-f_{M}g_{M}-\sum_{k=M}^{N}\Delta f_{k}g_{k}$
where $\Delta$ is the forward difference operator: $\Delta f_{k}=f_{k+1}-f_{k}$. %
} we can estimate $\sum_{m=M}^{\infty}\mu\left(m\right)m^{-\alpha}=O\left(M^{1/2-\alpha+\epsilon}\right)$
for $\alpha>1/2$. Plugging this estimate in Eq.~(\ref{eq:coeff_mu})
we obtain \begin{equation}
a_{n}^{L}-a_{n}\sim\frac{L^{-1/2-z+\epsilon}}{n^{1/2+\epsilon}}\,,\end{equation}
from which roughly $\sum_{n=1}^{L}\left(a_{n}^{L}-a_{n}\right)^{2}\sim L^{-1-2z+\epsilon}$. 

To summarize, in the case of analytic functions, relevant to massive
Left-Right symmetric models, we expect exponential convergence speed
with rate given by the correlation length. For critical theories with
dynamical exponent $z$, we expect an algebraic convergence speed.
Resorting to the Riemann Hypothesis we conjecture the convergence
speed to be of the order of $L^{-1-2z+\epsilon}$. 

\begin{figure}
\begin{centering}
\includegraphics[width=7cm]{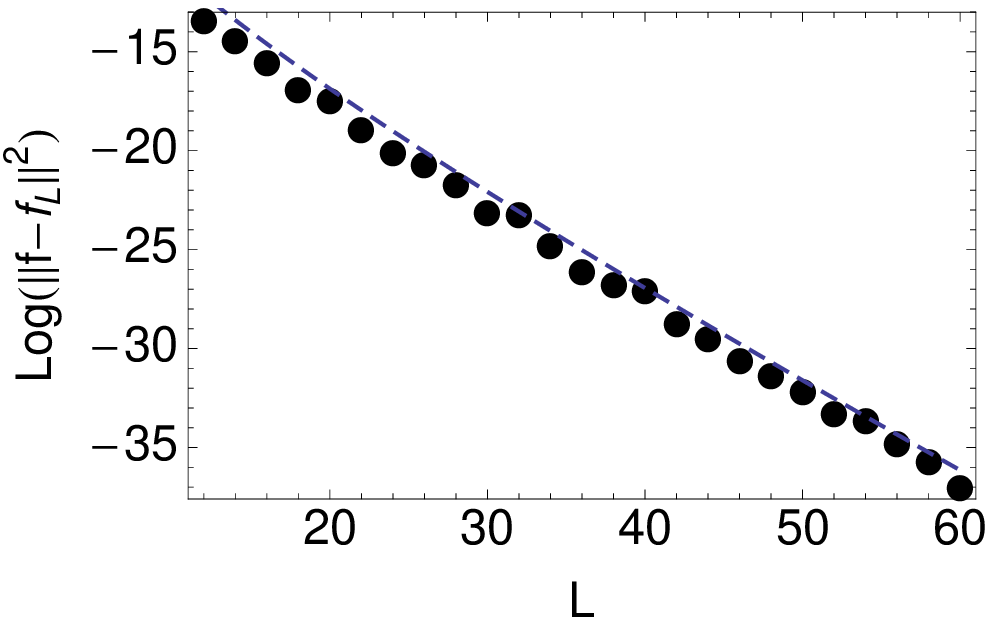}
\par\end{centering}

\begin{centering}
\includegraphics[width=7cm]{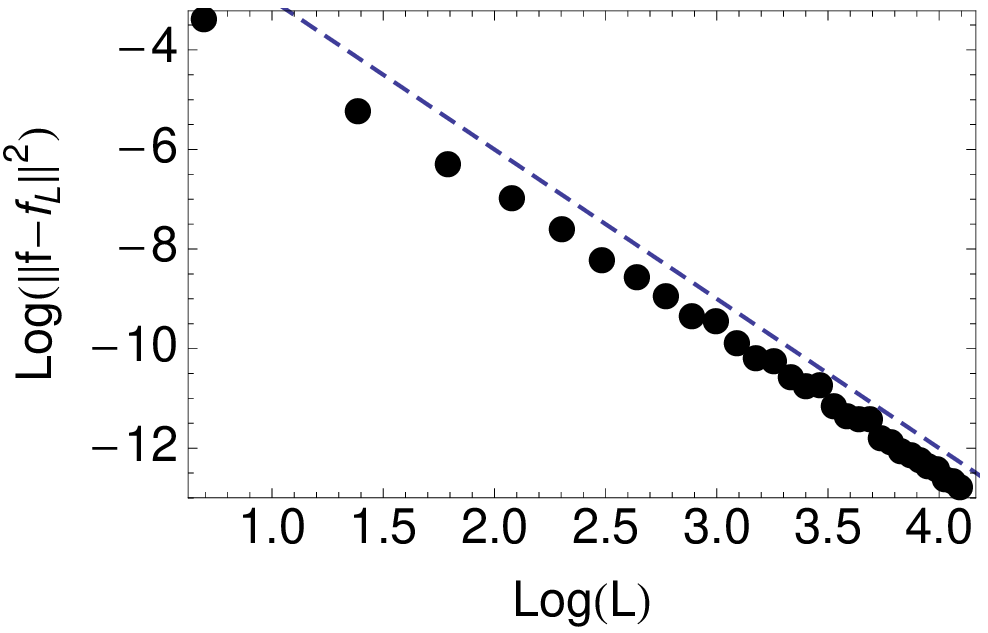}
\par\end{centering}

\caption{Convergence speed in the $L^{2}\left(\left[0,2\pi\right]\right)$
norm for the function $f\left(k\right)=\sqrt{\sin\left(k/2\right)^{2}+m^{2}}$.
The reconstructed function $f_{L}\left(k\right)$ is given by Eqns.~(\ref{eq:f_approx})
and (\ref{eq:finite_inverse}) and the $L^{2}\left(\left[0,2\pi\right]\right)$
distance $\left\Vert f-f_{L}\right\Vert _{2}^{2}$ is then evaluated
for various $L$. Top (bottom) picture refers to massive $m\neq0$
(massless $m=0$) case. In the massive case the approach expected
is of exponential type. The dashed line is given by $\mathrm{const.}\times L^{-3}e^{-2L/\xi_{E}}$,
with $\xi_{E}^{-1}=2\mathrm{arcsinh}\left(m\right)$ which can be
obtained estimating the Fourier coefficients of $f\left(k\right)$
with a saddle point argument. In the masless case the approach is
expected to be of the form $1/L^{3+\epsilon}$ (see main text). The
dashed line in the lower panel gives the $L^{-3}$ behavior. The method
is tested for PBC, and the mass in the top panel is $m=0.1$. \label{fig:norm_approach} }

\end{figure}

To visualize better how fast the method works, consider an example
taken from physics where the function to reconstruct is given by\begin{equation}
f\left(k\right)=J\sqrt{\sin\left(k/2\right)^{2}+m^{2}}.\label{eq:example}\end{equation}
Such a function is a periodic generalization of a relativistic dispersion
and it describes, exactly or approximately, the one-particle dispersion
of many one dimensional systems. When $m\neq0$ the function is analytic
in $\left[0,2\pi\right]$ and the convergence is exponentially fast.
The case $m=0$ can serve to model critical theories with dynamical
exponent $z=1$. In this case the convergence in the $L^{2}$-norm,
is algebraic and we just conjectured that the rate is of the order
of $L^{-3}$. This behavior is confirmed in figure \ref{fig:norm_approach}
which shows that, for the massive case, $\ln\left(\left\Vert f_{L}-f\right\Vert _{2}^{2}\right)$
is approximately linear with $L$, while for $m=0$ we have roughly
$\ln\left(\left\Vert f_{L}-f\right\Vert _{2}^{2}\right)\le-3\ln L+\mathrm{const.}$. 

\begin{figure}
\begin{centering}
\includegraphics[width=7cm]{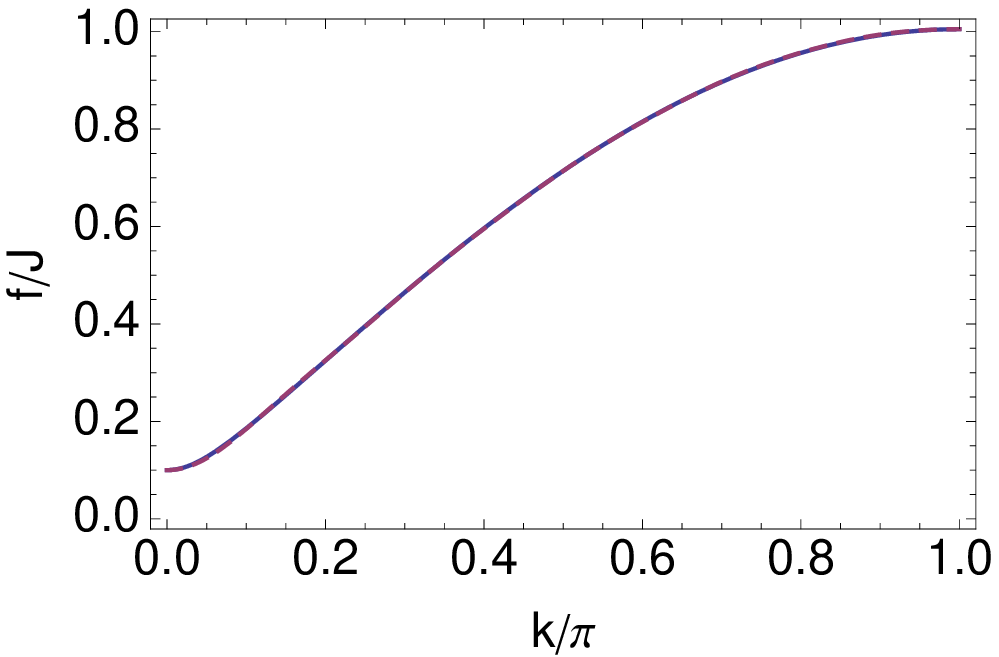}
\par\end{centering}

\begin{centering}
\includegraphics[width=7cm]{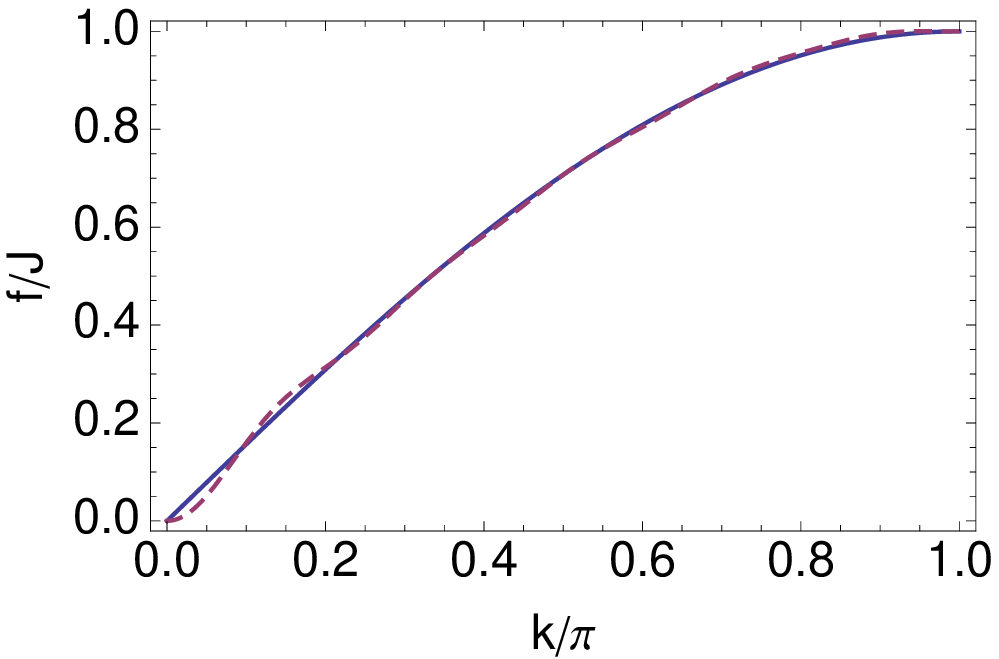}
\par\end{centering}

\caption{Results of the reconstruction for a prototypical function $f\left(k\right)=J\sqrt{\sin\left(k/2\right)^{2}+m^{2}}$,
using as little as ten Riemann sums. Upper panel refers to $m=0.1$
while lower panel to $m=0$. Continuous line is the exact function,
while the dashed line is the reconstructed function $f_{L}\left(k\right)$
using PBC. Since the functions are even only half of the Brillouin
zone is shown. Note that for $m=0.1$, $f_{L}\left(k\right)$ is indistinguishable
from the exact function at this scale. Results for ABC produce very
similar plots.  \label{fig:Result-of-the-approximation} }

\end{figure}

The result of the reconstruction for the function in Eq.~(\ref{eq:example})
is instead shown in figure \ref{fig:Result-of-the-approximation}
for different masses $m$, using PBC and as little as the first ten
Riemann sums. Results for ABC are very similar. It is notable the
very good agreement even in the massless case. 

\section{Method and applications\label{sec:Method-and-applications}}

The methods discussed so far are readily applicable to translation
invariant quasi-free systems consisting either of fermions or bosons.
By quasi-free systems we mean here Hamiltonians that can be expressed
as quadratic forms in Bose or Fermi operators. In such cases in fact
the ground state energy is precisely given by a partial Riemann sum.
For example, in the notation of \cite{LSM61} a quasi-free fermionic
model has the form $H=\sum_{ij}c_{i}^{\dagger}A_{i,j}c_{j}+1/2\left(c_{i}^{\dagger}B_{i,j}c_{j}^{\dagger}+\hc\right)$.
Diagonalization brings it to $H=\sum_{k}\omega_{k}\eta_{k}^{\dagger}\eta_{k}+\Gamma$,
where the band $\omega_{k}$ can be chosen positive and the constant
$\Gamma$ is given by $2\Gamma=\tr A-\sum_{k}\omega_{k}$. For translation
invariant systems, with PBC or ABC, the label $k$ is a (quasi-) momentum
quantized according to $k=\left(2\pi n+\vartheta\right)/L$, $n=0,1,\ldots,L-1$.
Defining the {}``filling fraction'' $\nu=1-\tr\left(A\right)/\sum_{k}\omega_{k}$
the Hamiltonian takes the form \begin{equation}
H=\sum_{k}\omega_{k}\left[\eta_{k}^{\dagger}\eta_{k}-\frac{\nu}{2}\right].\label{eq:quasi_free}\end{equation}
If the model consists of $n$ species of non-interacting colors, i.e.~$c_{i}\rightarrow c_{i,\alpha}$,~$\alpha=1,\ldots,n$,
simply replace $\nu\rightarrow n\nu$. All the methods presented so
far can now be applied considering that the ground state energy density
of Hamiltonian (\ref{eq:quasi_free}) is precisely given by $e_{L}=E_{L}/L=-\left(\nu/2\right)\sum_{k}\omega_{k}$.
Now the point is that in many physically interesting situations the
{}``filling fraction'' is known in advance. In fact in absence of
(magnetic or electric) fields generally $\nu=1$ (or $\nu=n$ for
$n$ non-interacting species) since $\tr A=0$. This means that the
$\nu$ in Eq.~(\ref{eq:quasi_free}) is independent of $L$, and
the ground state energy density is precisely given by a Riemann sum:
$e_{L}=-\left(\nu/2\right)S_{L}\left(\omega_{k}\right)$. Similar
considerations hold for bosonic quadratic theory with the important
difference that now $e_{L}=+\left(\nu/2\right)S_{L}\left(\omega_{k}\right)$,
due to the commutation relation. 

We can now argue that any interacting model admits some sort of quasi-free
approximation. At this level of approximation, the Hamiltonian is
quadratic and we can apply all the reasoning presented above. Our
method gives a way to obtain a one particle dispersion knowing the
ground state energy for some lattice sizes. The dispersion obtained
is optimal in the sense that it is the unique trigonometric polynomial
of degree $M$ (where $M$ is the number of energy data) consistent
with the observed values of the energy. This method bears some similarity
with the Hartree-Fock method largely used for ab-initio calculation
of molecular systems. The Hartree-Fock method, for a given size\emph{
}\inputencoding{latin1}{\emph{$L$}, gives the optimal quasi-free
state that minimizes the }\inputencoding{latin9}energy. The method
proposed here instead, taking $M$ ground state energies as input,
gives an optimal quasi-free system (identified with its one-particle
dispersion), in the sense that its ground state energies are precisely
the $M$ observed value. 

To specify completely the problem one has to assume the character
of the quasi-free approximation, i.e.~whether the model consists
of Bosons or Fermions together with the effective boundary conditions.
In practice we have to chose if the ground state energy densities
are given by $e_{L}=\epsilon\left(\nu/2\right)S_{L}\left(\omega_{k}\right)$
with $\epsilon=\pm1$ and moments specified by $\vartheta=0,\pi$
(PBC or ABC). This choice can be straightforward if the model under
consideration consists of Bosons or Fermions, but in case of spin
models, the character of the effective, quasi-free model is less clear.
According to the choices $\epsilon=\pm1$ and $\vartheta=0,\pi$ we
have therefore 4 possibilities. However the requirement that the reconstructed
band must be positive fixes in practice only two combinations. This
is an important result on its own: simply looking at the sequence
of ground state energies, one is able to assess whether quasi-particles
are Fermions or Bosons with ABC or PBC. 

As for any approximation method, it would be derisable to have a simple
criterion to assess whether a quasi-free approximation is feasible
in the first place. Such a criterion can be given. In fact for exactly
quasi free models, the ground state state energies satisfy\begin{equation}
E_{2L}^{\left(0\right)}=E_{L}^{\left(0\right)}+E_{L}^{\left(\pi\right)},\label{eq:criterion}\end{equation}
where the superscript $\left(0\right),\,\left(\pi\right)$ refers
to PBC, ABC respectively. So, having finite-size energies for PBC
and ABC, we can simply verify the possibility of an effective quasi-free
description by checking how well Eq.~(\ref{eq:criterion}) is satisfied. 

For what we have said in section \ref{sub:Covergence-rate}, the procedure
of reconstructing a function given its $M$ partial Riemann sums,
rapidly converges upon increasing $M$ even in the the critical case
(the worst scenario), so that one can effectively limit oneself to
short lattice sizes. 

For the reader's sake, let us sketch here the relevant steps of the
algorithm:
\begin{itemize}
\item Obtain a set of $M$ ground state energies of the system by exact
diagonalization of short lattices, say sizes up to $L=10-20$. If
both PBC and ABC energies are available one can check the feasibility
of a quasi-free approximation by checking how well Eq.~(\ref{eq:criterion})
is satisfied.
\item Assume effective PBC/ABC and Bosons/Fermions which correspond to assume
for the ground state energy density $e_{L}=\epsilon\left(\nu/2\right)S_{L}\left(\omega_{k}\right)$
with $\epsilon=\pm1$ and moments specified by $\vartheta=0,\pi$.
An approximate dispersion is then given by Eq.~(\ref{eq:f_approx})
with $f\left(k\right)=\epsilon\left(\nu/2\right)\omega_{k}$ and coefficients
specified by Eq.~(\ref{eq:finite_inverse}). The requirement $\omega\left(k\right)\ge0$
will fix two cases out of the four possibilities.
\item One should also fix the filling fraction $\nu$. If this can be simple
when the model is originally given in terms of Bosons or Fermions,
in general one must be guided by physical intuition. Referring to
the example that will be discussed in the following sections, it is
natural to expect $\nu=1$ for the pure spin-1/2 Heisenberg model,
a triplet of excitations for its dimerized version ($\nu=3$) and
a doublet ($\nu=2$) of excitations for the spin-1 $\lambda-D$ model
in the large-$D$ phase. 
\end{itemize}
Let us now illustrate how the method works on the hand of a few concrete,
yet prototypical examples.

\subsection{Spin 1/2 Heisenberg model}

\begin{figure}
\begin{centering}
\includegraphics[width=8cm]{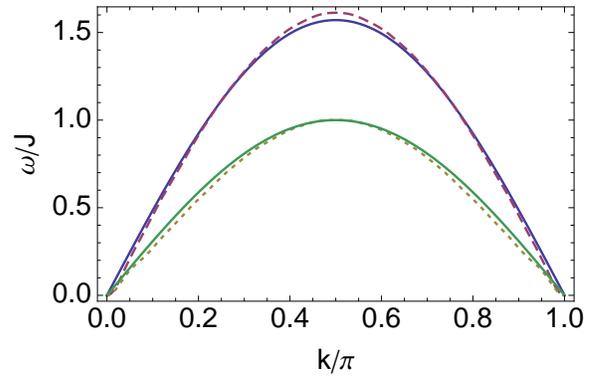}
\par\end{centering}

\caption{(Color online) Results of our procedures for the spin-1/2 Heisenberg
model. The continuous upper curve (blue) is the exact dispersion $\omega\left(k\right)=$$\left(\pi/2\right)J\left|\sin\left(k\right)\right|$,
while the continuous lower curve (green) is the result of spin-wave
approximation $\omega_{SW}\left(k\right)=J\left|\sin\left(k\right)\right|$.
Using only finite size energies up to $L=24$ we obtained upper and
lower dashed curve. The lower curve is obtained assuming a bosonic
dispersion with effective PBC and can be identified with the spin-wave
approximation. The upper dashed curve assumes a dispersion with fermionic
character and effective ABC. The very good agreement with the exact
result tells us that there exist a quadratic Hamiltonian approximating
Heisenberg one very precisely. \label{fig:heisenberg} }

\end{figure}

Take the Heisenberg antiferromagnetic ($J>0$) chain: \begin{equation}
H=J\sum_{i=1}^{L}\mathbf{S}_{i}\cdot\mathbf{S}_{i+1}\,.\label{eq:heisenberg_ham}\end{equation}
$\mathbf{S}_{i}$ are spin-1/2 operators at site $i$ and PBC ($\mathbf{S}_{L+1}=\mathbf{S}_{1}$)
are used. From the exact solution we know that the infinite size ground
state energy is $e_{\infty}/J=1/4-\ln2$ \cite{hulthen38}, whereas
the quasi-particle dispersion is given by $\omega\left(k\right)=\left(\pi/2\right)J\left|\sin\left(k\right)\right|$
\cite{descloizeaux62}. When we have to evaluate the energy at finite
size we immediately face a problem. When $L$ is even the ground state
belongs to the total spin $S=0$ sector and is unique \cite{liebmattis62}.
This is the kind of ground state we {}``expect'' from this model.
On the contrary, for $L$ odd there are two degenerate spin-1/2 ground
states. The ground state energies for $L$ odd have a completely different
character and our intuition suggests us to discard them. As a consequence
we have access only to $e_{L}$ for even $L$. However, from the exact
solution we know that the dispersion is periodic with period halved
i.e.~$\pi$. Hence it has only even Fourier (cosine) coefficients.
From Eq.~(\ref{eq:finite_inverse}) we see that with even-size energies
we can re-construct even Fourier coefficients. In this case the two
facts are consistent: $e_{L}$ only for $L$ even $\leftrightarrow$
even Fourier coefficients. The procedure is as follows. First, we
diagonalize exactly Hamiltonian (\ref{eq:heisenberg_ham}) with say
a Lanczos algorithm. In few minutes of a small laptop computer, we
obtained ground state energies for lattices of even size up to $L=24$
. Separately we estimate the infinite size ground state energy which
in this case is $e_{\infty}=J(1/4-\ln2)$. Having collected the numbers
$R_{L}$ for $L$ even, we can use Eq.~(\ref{eq:finite_inverse})
to obtain a one-particle dispersion. To specify the problem completely
we have to make few further assumptions. First we have to fix boundary
conditions. Even if we have PBC for the spins different BC's can be
induced in the effective quasi-free model. Indeed using the Jordan-Wigner
transformation, model (\ref{eq:heisenberg_ham}) can be exactly mapped
to a model of interacting spinless fermion with parity dependent boundary
conditions (see for example \cite{lcv10} for a discussion on these
emerging BCs). Since the ground state is a singlet it belongs to the
parity one sector, where BC's for the fermions are anti-periodic.
So, to be more general, we consider equation (\ref{eq:finite_inverse})
for $q=1,\,-1$ which corresponds to effective PBC or ABC respectively.
On physical grounds %
\footnote{For example in the spin-wave approximation of Anderson and Kubo \cite{anderson52,kubo53}
one would identify quasiparticles as spinless bosons, while using
the Jordan-Wigner transformation one would conjecture spinless fermions
(see also below). In both cases there is only one copy of bosons or
fermions, i.e.~$\nu=1$. %
} we fix the filling fraction to $\nu=1$. This is enough to obtain
a $q$ dependent function $f\left(k\right)=\left(\epsilon/2\right)\omega_{k}$.
To specify completely the dispersion we must still decide whether
the effective quasi-particles are either fermions ($\epsilon=-1$
) or bosons ($\epsilon=+1$). The four possible cases corresponding
to $q=\pm1$ and $\epsilon=\pm1$ are reduced to two by imposing positivity
of the band. The result is that assuming effective PBC quasi-particles
are Bosons, while assuming ABC quasi-particles must be Fermions.

The results of the procedure, using only PBC ground state energies
up to $L=24$, are shown in Fig.~(\ref{fig:heisenberg}). The bosonic
dispersion with effective PBC can be identified with the spin-wave
dispersion $\omega_{SW}\left(k\right)=J\left|\sin\left(k\right)\right|$,
obtained with the spin-wave approximation of Anderson and Kubo \cite{anderson52,kubo53}
{[}see lower curves in Fig.~(\ref{fig:heisenberg}){]}. Instead,
the Fermionic dispersion with effective ABC is very close to the exact
one of des Cloizeaux and Pearson. The excellent agreement of this
dispersion with the exact one, indicates that a good description (as
long as short range quantities are concerned) of the Heisenberg model
can be given in terms of an effective quasi-free fermionic Hamiltonian
with ABC. 

Although the exact one-particle dispersion of the spin-1/2 Heisenberg
model could be reproduced with high precision, this example also shows
a limitation of our method. Namely to use our method we need either
ground state energies $e_{L}$ for general $L$, or if we only have
access to even sizes $L$ we can only reconstruct even Fourier coefficients.
These limitations disappears if we consider dimerized models where
we expect the dispersion to be $\pi$-periodic. This is because an
even function of period $\pi$ has only even (cosine) Fourier coefficients.
Knowing finite size energies for even sizes $L$ is enough to reconstruct
--within a certain approximation-- the whole one-particle dispersion.

\subsection{Dimerized spin-1/2 chain}

Consider then a spin-1/2 Heisenberg model with an explicit dimerization
of the exchange coupling:\begin{equation}
H=J\sum_{i=1}^{L}\left[1+\delta\left(-1\right)^{i}\right]\mathbf{S}_{i}\cdot\mathbf{S}_{i+1}\,.\label{eq:ham_dimer}\end{equation}

\begin{figure}
\begin{centering}
\includegraphics[width=8cm]{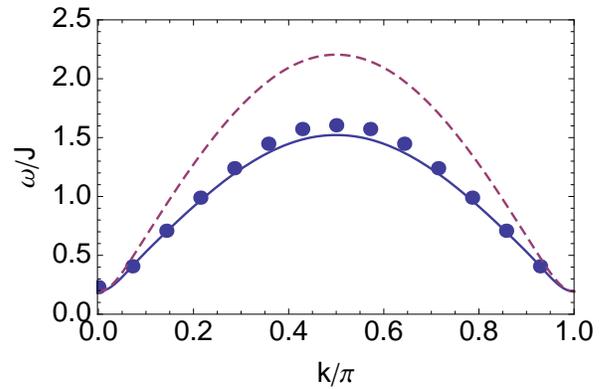}
\par\end{centering}

\caption{One particle dispersion of model (\ref{eq:ham_dimer}) at $\delta=0.048$.
The full dots are exact diagonalization data of \cite{stephan-spin-peierl}
obtained for a chain of $L=28$ sites (reproduction with permission
of the authors). Our method is tested using only finite size energies
up to $L=12$. The dashed curve is the result of our method assuming
a triplet of fermions with effective ABC, while the continuous curve
assumes a triplet of bosons with effective PBC. \label{fig:dimer_model} }

\end{figure}

This model has been extensively used to characterize a variety of
spin-Peierls compounds. The presence of the dimerization has the effect
of halving the Brillouin zone and so, correspondingly, the one-particle
dispersion should have period $\pi$. That this is indeed the case
is confirmed by many numerical simulation \cite{stephan-spin-peierl}.
Then we can safely use even size energies to reconstruct the even
Fourier coefficients of the dispersion. Moreover, a non-zero $\delta$
has also the effect of opening a mass gap. As discussed in section
\ref{sub:Covergence-rate}, the convergence rate of our method is
expected to be extremely fast in this case. With the aim of showing
the usefulness of the method, we consider very short length. Using
only finite size energies at even sizes from $L=2$ to $L=12$ (i.e.~only
six numbers!) we obtain the dispersion shown in Fig.~\ref{fig:dimer_model}.
The results are then compared with those obtained via much more powerful
diagonalization of ref.~\cite{stephan-spin-peierl} performed on
a chain of $L=28$ sites. 

\subsection{Spin-1 model with single ion anisotropy}

\begin{figure}
\begin{centering}
\includegraphics[width=8cm]{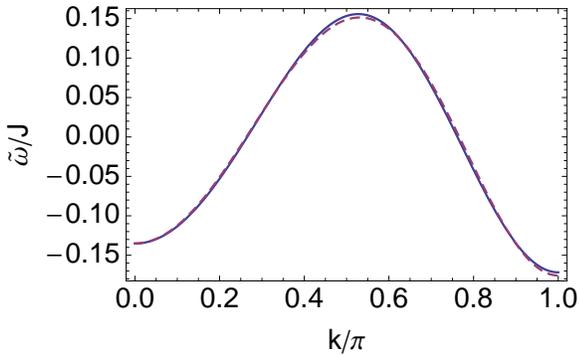}
\par\end{centering}

\caption{One particle dispersion for model Eq.~(\ref{eq:ham-D}) in the large-$D$
phase ($D=7.4$). Only the part of the function corresponding to Fourier
cosine coefficients $c_{n}$ $n>1$ are shown. The continuous curve
is the result of third order perturbation theory Eq.~(\ref{eq:omega-3})
without the $\cos\left(k\right)$ term. Perturbation theory is basically
exact for such value of $D$ according to \cite{golinelli-D92}. Our
procedure is tested using finite size energies (obtained by Lanczos
method) of short chains from $L=2,3,\ldots$up to $L=12$ and later
discarding the $\cos\left(k\right)$ term from the dispersion. The
dashed curve is our result assuming the dispersion is a doublet of
bosonic particle and effective PBC. Assuming a doublet of Fermi particles
and ABC one obtains similar accuracy although the $L^{2}\left(\left[0,\pi\right]\right)$
norm distance with dispersion (\ref{eq:omega-3}) favors the bosons
(distance $0.049$ against $0.078$). \label{fig:large-D} }

\end{figure}

As we have shown, our methods can be successfully applied to spin-1/2
chains only when the dispersion is even and of period $\pi$. This
is the case for the pure Heisenberg model and for dimerized models
as the one of Eq.~(\ref{eq:ham_dimer}). What about spin-1 chains?
For PBC and even size the theorem by Lieb and Mattis \cite{liebmattis62}
tells us that the ground state of a generic antiferromagnetic Heisenberg
model belongs to the total spin zero sector and is unique. For odd
sizes an antiferromagnet with PBC is frustrated and the theorem does
not apply. However we have numerically verified that also for odd
sizes the ground state belongs to the total spin zero sector (this
is consistent with the VBS description and with the fact that every
spin-1 can be thought of a symmetric combination of two spin-1/2,
and so any chain contains an even number of spin-1/2). This suggests
that we could use ground state energies both for even and odd sizes
and re-construct completely the dispersion. However there is still
a problem with this approach. The ground state energy for the single
site problem is not clearly defined. If the model admits a quasi-free
approximation, the $L=1$ ground state energy is given by $E_{1}=\pm\left(\nu/2\right)S_{1}\left(\omega_{k}\right)=\pm\left(\nu/2\right)\omega_{\vartheta}$
(plus or minus refers to Bosons or Fermion respectively). Using the
inversion Eq.~(\ref{eq:finite_inverse}) we see that, $S_{1}\left(f\right)$
enters only in the definition of the first Fourier coefficient $a_{1}$.
So a missing $S_{1}\left(f\right)$ allows to specify the function
up to an additive $\cos\left(k\right)$ term. This term could be fixed
by other means, such as obtaining the value of the dispersion at a
given momentum. Let us analyze a concrete example.

Consider the spin-1 model with single ion anisotropy\begin{equation}
H=J\sum_{i=1}^{L}\left[\mathbf{S}_{i}\cdot\mathbf{S}_{i+1}+D\left(S_{i}^{z}\right)^{2}\right]\,.\label{eq:ham-D}\end{equation}
where $\mathbf{S}_{i}$ are now spin-1 operator. In order to test
our method we consider the model for large $D$ where perturbation
theory is applicable and an analytic expression for the dispersion
is available. When $D=\infty$ the ground state is given by $|0,0,\ldots,0\rangle$,
and excitations form a doublet of degenerate states with the spin
at one site flipped to $+1$ or $-1$, i.e.~$|0,0,\ldots,\pm1,\ldots,0\rangle$.
A finite large $D$, removes translation degeneracy and one obtains
a doubly degenerate band $\omega\left(k\right)$. This picture remains
valid in the whole, so-called, large-$D$ phase, which is separated
by the Haldane phase roughly at $D_{c}\simeq1$ (see \cite{chenhidasanctuary03}
for details on the phase diagram). In the large-$D$ phase one can
use perturbation theory to obtain the doubly degenerate dispersion.
A third order calculation has been performed \cite{papanicolau90}
with the result 

\begin{multline}
\frac{\omega\left(k\right)}{J}=D+2\cos\left(k\right)+\frac{1}{D}\left[1+2\sin\left(k\right)^{2}\right]+\\
\frac{1}{D^{2}}\left[2\sin\left(k\right)^{2}-\frac{1}{2}(1+8\sin\left(k\right)^{2})\cos\left(k\right)\right]+O\left(D^{-3}\right)\label{eq:omega-3}\end{multline}
 We re-write the dispersion as \begin{multline}
\frac{\omega\left(k\right)}{J}=\left(D+\frac{2}{D}+\frac{1}{D^{2}}\right)+2\left(1-\frac{3}{4D^{2}}\right)\cos\left(k\right)\\
-\frac{1+D}{D^{2}}\cos\left(2k\right)+\frac{1}{D^{2}}\cos\left(3k\right)+O\left(D^{-3}\right)\,,\label{eq:omega-3b}\end{multline}
in order to make clear the Fourier (cosine) coefficients of the dispersion.
Using a Lanczos algorithm we computed the ground state energy of the
model (\ref{eq:ham-D}) for $L=2,3,\ldots,12$. For what we have said,
using the inversion Eq.~(\ref{eq:finite_inverse}) we can reconstruct
the band up to a cosine term. In figure \ref{fig:large-D} we show
the result for the reconstructed band compared to the dispersion Eq.~(\ref{eq:omega-3b})
both without the $\cos\left(k\right)$ term and the agreement is excellent.
As noticed previously, the $\cos\left(k\right)$ term can be fixed
by other means.

\section{Conclusions}

In this article we showed that a lot more information than currently
believed, is encoded in the ground state energy density at finite
size $e_{L}$. In particular we provided a method able to reconstruct
an approximate one-particle dispersion for any one-dimensional quantum
system, given some finite size numerical data $\left\{ e_{L}\right\} $.
The dispersion reconstructed with this procedure is optimal in the
sense that it is the unique trigonometric polynomial of degree $L_{\mathrm{max}}$
($L_{\mathrm{max}}$ being the number of energy data) consistent with
the observed data $\left\{ e_{L}\right\} $. Equivalently the method
produces a quasi-free Hamiltonian which has the same ground state
energy densities as the observed values $\left\{ e_{L}\right\} $.
This method is exact if the model has some sort of quasi-free representation,
and it converges very rapidly increasing $L_{\mathrm{max}}$ so that
very few data are sufficient (using 10 energy data gives already very
good results). We also provided a simple criterion to assess whether
such a quasi-free approximation is feasible in the first place. As
a side effect, simply looking at the sequence $\left\{ e_{L}\right\} $
this method is able to assess whether effective quasiparticles are
either boson or fermions with effective periodic or anti-periodic
boundary conditions. 

Since the Casimir force is specified (up to a constant) by the energies
$\left\{ e_{L}\right\} $, from a physical point of view the procedure
presented consists on reconstructing the one-particle dispersion given
the Casimir force. 

Further developments in this direction include the possibility of
extending these ideas to higher dimension and testing the procedure
on other strongly correlated systems. 
\begin{acknowledgments}
The author would like to thank M.~Roncaglia for useful discussions
on the spin-1 model. 
\end{acknowledgments}
\bibliographystyle{apsrev}
\bibliography{nrg}

\end{document}